\documentclass[journal]{IEEEtran}
\usepackage{myStyleIEEE}
\usepackage{ushort}
\usepackage{threeparttable}
\usepackage{makecell}
\usepackage{lettrine}
\usepackage{dblfloatfix}

\IEEEoverridecommandlockouts

\usepackage{siunitx}
\AtBeginDocument{\DeclareSIUnit{\MWh}{MWh}}
\AtBeginDocument{\DeclareSIUnit{\kWh}{kWh}}
\DeclareSIUnit \voltampere { VA } 
\DeclareSIUnit \var { var } 
\usepackage{amsmath}
\usepackage{mathtools}
\usepackage{isomath}

\usepackage{fancyhdr}
\usepackage{amsmath,amsfonts,bm,mathtools}

\def\BibTeX{{\rm B\kern-.05em{\sc i\kern-.025em b}\kern-.08em
    T\kern-.1667em\lower.7ex\hbox{E}\kern-.125emX}}

\usepackage{titlesec}
\begin{document}

\title{Improved Droop Control in DC Microgrids via Voltage-Locked Loop Synchronization}
\renewcommand{\theenumi}{\alph{enumi}}

\author{
\IEEEauthorblockN{Ognjen Stanojev,~\IEEEmembership{Member,~IEEE,} Jovan Krajacic,~\IEEEmembership{Student~Member,~IEEE,} Enea Bianda,\\ Orcun Karaca,~\IEEEmembership{Senior~Member,~IEEE,} Mario Schweizer,~\IEEEmembership{Senior~Member,~IEEE}}%
\thanks{O. Stanojev, E. Bianda, O. Karaca and M. Schweizer are with ABB Corporate Research Center, Switzerland, emails:\{ognjen.stanojev, enea.bianda, orcun.karaca, mario.schweizer\}@ch.abb.com}
\thanks{J. Krajacic is with Power Systems Laboratory at ETH Z\"{u}rich, CH-8092,  Switzerland, email:\{jkrajacic\}@eeh.ee.ethz.ch.}
}

\maketitle
\thispagestyle{fancy}
\IEEEpeerreviewmaketitle

\begin{abstract}
DC microgrids are low- or medium-voltage networks designed to connect and manage DC-based sources and loads.
A key challenge in operating DC microgrids is maintaining the DC voltage within certain
predefined limits while ensuring its stability. Droop control, the most common method towards addressing this challenge, enables decentralized voltage control and power sharing, but suffers from poor transient performance, resulting in voltage dips and overshoots in applications with fast varying loads such as AI datacenters.
This paper introduces an improved droop control method based on voltage-locked loop synchronization, which ensures DC voltage stabilization with significantly
improved transient response and achieves the desired load sharing between the available source
converters. The core design principle is reflected in the functional separation of the control scheme into a virtual DC machine (VDCM) that operates as a spinning wheel and a virtual current source connected in parallel. The transient response is provided by the VDCM to damp and stabilize DC bus voltage variations, while the slow droop response is provided by the virtual current source to ensure steady-state power balance. The performance of the proposed method is validated 
in an experimental microgrid setup.
\end{abstract}

\begin{IEEEkeywords}
DC microgrids, datacenters, droop control, transient response, virtual DC machine, voltage-locked loop.
\end{IEEEkeywords}

\section{Introduction} \label{sec:intro}
DC microgrids are emerging as an effective solution to integrate energy resources that naturally operate at DC voltage, such as batteries and photovoltaic systems, and connect them directly to local DC consumption, such as electric vehicles and datacenter IT loads. Similarly to the role of kinetic energy stored in rotating machines in AC systems, stored energy in capacitances in the DC network is used as an energy buffer between supply and demand \cite{Krajacic2026}. Therefore, DC voltage is an indicator of mismatch between load and supply and must be controlled within tight limits, e.g., $320$-$380\,\mathrm{V}$ for the nominal voltage of $350\,\mathrm{V}$, as defined in the Current/OS standard~\cite{currentos2025}.

The most widely considered control technique in both AC and DC microgrids is droop control \cite{Guerrero2011}. It is a decentralized control strategy implemented in the form of a proportional controller, allowing power sharing among parallel units without the need for communication \cite{Meng2017}. In DC microgrids, virtual resistance-based droop control \cite{Tahim2015} is typically implemented for DC voltage control, where virtual resistance is defined as the maximum allowed voltage deviation divided by the nominal converter output current. Thus, the load is shared among the source converters in proportion to their power (current) ratings, and the DC voltage remains within the allowable range. 

Two main types of droop control have emerged~\cite{Gao2017,CompareDroop2017,Wang2018}: current mode (I--V) and voltage mode (V--I).
In current mode, the output voltage error is divided by the virtual resistance to obtain a current reference for the inner current controller, whereas in voltage control mode, the voltage drop arising from virtual resistance multiplied by the output current is subtracted from the DC voltage setpoint to generate a voltage reference for the inner voltage controller. The former method features a straightforward control structure and is easy to implement.  However, without an additional low-pass or lead-lag filter for voltage error processing, stability can be compromised for droop values required for a tight voltage band with high-power converters~\cite{Gao2017}. Unfortunately, the added filter then degrades transient performance. The latter method offers an improvement in both the stability margin and the transient response~\cite{CompareDroop2017}, but has a dual-loop control structure that might be difficult to tune and requires an additional (current) measurement.

Numerous works have focused on improving the performance of droop control~\cite{Tomislav2016}. These approaches can be broadly classified into five categories: adaptive droop techniques~\cite{Augustine2015,Jiang2021,Zhao2024}, nonlinear droop control~\cite{Prabhakaran2018,Chen2019,Yang2024}, communication-based methods~\cite{Anand2013,Lu2014,Wang2016}, output impedance shaping methods~\cite{Zhu2020,RCmode2017,Deng2024,Liu2020}, and virtual DC machine (VDCM) control~\cite{VDCM2016,VDCM_SOC2020,PowerLoopFree2022,Gang2021}. The adaptive droop control techniques aim to achieve more accurate load sharing within a specified range of bus voltage deviation. However, practical implementation of this control strategy is hindered by the complexity of transitioning between different segments of the droop curves, which can induce undesirable transients and pose potential stability issues~\cite{Zhao2024}. Similarly, communication-based methods can improve current sharing and achieve more optimal microgrid operation, but require a reliable communication infrastructure and can be affected by delays \cite{Wang2016}. Nonlinear droop control methods can provide a faster transient response compared to canonical dual-loop droop control due to the elimination of external DC voltage or current loops \cite{Yang2024}. However, the introduced nonlinearities complicate control design and network analysis.

Output impedance shaping methods aim to enhance droop-based DC voltage regulation by shaping the converter's output impedance such that it is purely resistive at medium and low frequencies, and capacitive at high frequencies~\cite{Liu2020}. In this way, the transient current is mainly provided by the converter, and the current flow through the filter capacitor is minimized \cite{Deng2024,Liu2020}. In \cite{Deng2024}, it is demonstrated that the equivalent output impedance of the converter can be significantly reduced in magnitude in the medium frequency range by controlling the current of the filter capacitor rather than the current of the filter inductor, and by adding a virtual capacitance at the converter's output. The I-V droop performance is improved in \cite{Zhu2020} by adding a virtual capacitor at the converter output and by considering an additional damping current component. Similarly, the work in \cite{RCmode2017} proposes an RC-mode droop control to improve the transient response of both I-V and V-I modes. Nevertheless, impedance shaping lacks generality, as it is confined to specific converter topologies with LC output filters. Moreover, the resulting controllers often lack physical interpretability.\looseness=-1

Drawing inspiration from virtual synchronous machine control \cite{VSM2007} designed for grid-forming converters in AC grids, the concept of VDCM control has been proposed for converters in DC microgrids \cite{VDCM2016,VDCM_SOC2020,PowerLoopFree2022,Gang2021}. This concept was first explored in \cite{VDCM2016}, where an input-output relationship between mechanical power input and armature current is used to emulate the DC machine. A voltage controller is added to adjust the mechanical power of the virtual DC machine, thereby stabilizing the DC bus voltage. 
In \cite{PowerLoopFree2022}, the VDCM control is simplified by eliminating the need to calculate torque from power measurements. However, an additional voltage controller is still required, and its bandwidth is limited by the dynamics of the VDCM. 
Furthermore, VDCM approaches introduce excessive complexity into the control structure due to the nonlinear relationships associated with power and torque calculations and rely on a large number of parameters.

In this paper, we present a novel converter control scheme for DC microgrids that emulates the joint effect of a VDCM and a virtual controlled current source connected in parallel. This functional separation enables a simple and modular control design. Furthermore, the functional separation corresponds to the timescale separation of the converter response, where the transient response is provided by the VDCM to damp and stabilize DC bus voltage variations, while the slow response is provided by the virtual current source droop to ensure steady-state power balance. 

The VDCM part of the control scheme is emulated by combining a voltage-locked loop (VLL) with a virtual admittance. Based on the DC machine model (with no mechanical input power), a relationship between the voltage applied to the machine terminals and the electromotive force (EMF) voltage generated by the spinning wheel can be established. We demonstrate that this transfer function behaves as a VLL and has a strong resemblance to the phase-locked loop (PLL) used in AC grid-connected converters~\cite{SyncReview2020}. To the best of our knowledge, this paper is the first to propose such a feature. 

The virtual current source provides droop control by injecting current proportionally to the mismatch between the virtual EMF voltage and the desired voltage setpoint. With this design of the virtual current source, the response that resembles the governor action in AC systems is achieved. In contrast to basic droop control approaches, virtual EMF voltage is used for droop control instead of the direct output voltage measurement, resulting in a smoother dynamic response and enhanced stability characteristics. Unlike VDCM methods, the droop output (reference) is directly passed to the current controller without filtering it with the DC machine dynamics, thus improving setpoint tracking performance. 

By combining the two above-described functionalities, an improved droop control method based on VLL synchronization is obtained, which is hereafter referred to as VLL-based improved droop control (VLL-IDC).
The two functionalities create separate current references that can be impressed by the converter using any current controller variant used in DC microgrids. Therefore, the proposed method can seamlessly be integrated in the existing converter controls. Furthermore, the \textit{amount} of transient response and the \textit{amount} of steady-state power injection can be tuned independently, i.e., it is possible to configure the scheme to provide only transient response or only steady-state power sharing. The parameter that determines the steady-state contribution is the droop gain, whereas the parameter that defines the transient contribution is the virtual admittance. This flexibility and modularity is unique in the context of the state-of-the-art DC microgrid control approaches. It allows, e.g., a converter that is in a constant power section of a piecewise droop curve, to provide transient bus voltage stabilization. 

In summary, the main novel features of the proposed scheme, considering the requirement to regulate the DC bus voltage with acceptable transients and steady-state deviations, and compared to the previous schemes in the literature, are:
\begin{itemize}
	\item Seamless droop control: the method is compatible with upcoming DC microgrid standards~\cite{currentos2025} and guarantees a seamless droop curve on the DC output. This property is provided by the virtual current source functionality.
	\item Improved transient performance: the method offers enhanced transient performance with appropriate damping compared to standard droop control techniques. This property is provided by the virtual DC machine. It allows meeting stringent voltage limits specified by DC standards during transients and avoids the need to install excessive capacitance for the DC bus.
	\item It has the possibility to configure the amount of transient and steady-state support and the capability to provide transient grid support and improve DC bus voltage quality even when the converter is in a region of constant power defined by a piecewise droop curve, e.g., with a battery or with insensitive loads.
	\item Only the DC bus voltage measurement is required. This is in contrast to previous work on droop control methods and their derivatives, which often require additional measurements such as output or capacitor current.
	\item It has a simple control structure, in contrast to previous VDCM methods, which employ complicated nonlinear control structures. Moreover, a small number of parameters need to be tuned. Furthermore, it is compatible with existing current controllers. These features imply a low commissioning effort.
\end{itemize}

It is important to note that the proposed method follows the same
core principles as grid-forming vector current control~\cite{Schweizer2022}, recently developed for AC grid-connected converters, and can be considered its DC counterpart. Furthermore, despite the usage of a VLL, the controller does not rely on an energized DC bus, i.e., it is fully grid-forming~\cite{DCGridForming2026} and can operate in stand-alone or black start an unenergized DC bus.

The rest of the paper is structured as follows. Section~\ref{sec:vll} introduces the voltage-locked loop synchronization principle and its core components. Section~\ref{sec:GFVCC_overview} presents in detail the VLL-based improved droop control method.
Section~\ref{sec:outputimpedance} provides small-signal analysis and output impedance characterization with parameter tuning guidelines.
Section~\ref{sec:exp} validates the proposed method experimentally under various operating conditions and compares it with conventional droop control. Finally, Section~\ref{sec:concl} concludes the paper.

\section{Voltage-Locked Loop}\label{sec:vll}
Similarly to the phase-locked loop employed in the control of AC grid-connected converters~\cite{SyncReview2020}, a voltage-locked loop can be formulated for the control of DC grid-connected converters. In contrast to the AC case, where synchronization is required to track the grid voltage phase, DC converter systems require accurate regulation and alignment with the DC bus voltage (magnitude) to ensure stability and precise power sharing. The VLL synchronization thus provides a systematic framework for processing the DC bus voltage, thereby enhancing disturbance rejection and supporting stable converter operation under varying conditions. Since the applicability of this concept extends beyond the proposed control strategy, it is presented here in a separate, dedicated section.

\begin{figure}[b]
	\centering
	\includegraphics[scale=1]{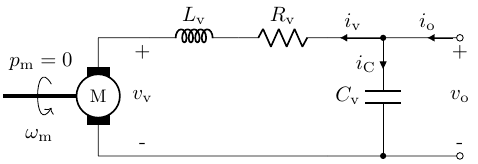}
	\caption{Virtual DC machine representation.}
	\label{fig:virt_DCm}
\end{figure}

Derivation of the VLL is based on the circuit in Fig.~\ref{fig:virt_DCm}, where a DC machine is supplied from a DC bus of voltage $v_\mathrm{o}$. A virtual capacitor $C_\mathrm{v}$ is considered at the DC-bus. Armature circuit is characterized by resistance $R_\mathrm{v}$ and inductance $L_\mathrm{v}$, and $v_\mathrm{v}$ represents the electromotive force generated in the armature. The machine operates without any mechanical input power applied to its shaft, at the rotor speed $\omega_\mathrm{m}$. 

The basic equations for static analysis of DC machines describe the EMF generation process and its relationship to the terminal (output) voltage \cite{Fitzgerald1990}:
\begin{subequations}
	\begin{align}
		v_\mathrm{v} &= K_\mathrm{a}\Phi_\mathrm{d}\omega_\mathrm{m}, \\
		v_\mathrm{o} &= v_\mathrm{v} + R_\mathrm{v} i_\mathrm{v}, \label{eq:armature_drop}
	\end{align}
\end{subequations}
with $K_\mathrm{a}$ denoting the winding constant and $\Phi_\mathrm{d}$ is the direct-axis air-gap flux per pole. For brevity, the effect of these constants is aggregated in $K_\omega=K_\mathrm{a}\Phi_\mathrm{d}$. Under small-signal assumptions (linearization around nominal angular rotor speed $\omega_\mathrm{n}$), the equation of motion for the rotor can be expressed in the following form (note that both mechanical and electrical inputs are directed \textit{towards} the machine):
\begin{equation}\label{eq:swing_eq_dcm}
	\omega_\mathrm{n}J_\mathrm{v}\ddt{\omega_\mathrm{m}} = p_\mathrm{m} + p_\mathrm{e},
\end{equation}
where $J_\mathrm{v}$ is the total (virtual) moment of inertia. As already established, the mechanical input is not connected, i.e., $p_\mathrm{m}=0$. Due to negligible losses in the armature circuit, we approximate the electrical power as $p_\mathrm{e}\approx p_\mathrm{o}=v_\mathrm{o}i_\mathrm{o}$. Furthermore, $p_\mathrm{o}$ is linearized by assuming that the change in power primarily results from the change in current and that the output voltage is regulated within tight bounds. Using \eqref{eq:armature_drop}, the output power can be expressed in the Laplace domain as \begin{equation}\label{eq:output_power} p_\mathrm{o} = i_\mathrm{o}V_\mathrm{o} = (i_\mathrm{v}+i_\mathrm{C})V_\mathrm{o} = \left(\frac{v_\mathrm{o}-v_\mathrm{v}}{R_\mathrm{v}} + sC_\mathrm{v}v_\mathrm{o}\right)V_\mathrm{o}. \end{equation}
Recall that capital letters, in this case $V_\mathrm{o}$, indicate that the quantity is constant. This expression is derived under the quasi steady-state assumption introduced previously, where the inductive voltage drop $sL_\mathrm{v}i_\mathrm{v}$ is neglected with respect to the resistive drop $R_\mathrm{v}i_\mathrm{v}$. Physically, this is justified since the electrical transient associated with the armature inductance evolves on a much faster time scale than the dynamics of interest, so the current remains close to its steady-state value, $i_\mathrm{v} \approx {(v_\mathrm{o}-v_\mathrm{v})}/{R_\mathrm{v}}$. In contrast, the capacitive contribution $sC_\mathrm{v}v_\mathrm{o}$ is retained since the output capacitance can store a significant amount of energy and introduces dynamics on time scales comparable to the quasi steady-state behavior under consideration. Transforming \eqref{eq:swing_eq_dcm} to Laplace domain and combining it with \eqref{eq:output_power}, one obtains 
\begin{align}
	\omega_\mathrm{n}J_\mathrm{v}\,s\omega_\mathrm{m} &= p_\mathrm{e} \approx p_\mathrm{o}, \\
	\omega_\mathrm{n}J_\mathrm{v}\,sK_\omega^{-1} v_\mathrm{v} &= (\frac{v_\mathrm{o}-v_\mathrm{v}}{R_\mathrm{v}}+sC_\mathrm{v}v_\mathrm{o})V_\mathrm{o}.
\end{align}
An input-output relationship between the output voltage $v_\mathrm{o}$ and the virtual back EMF voltage $v_\mathrm{v}$ can be established by assuming $\omega_\mathrm{n}\approx K_\omega V_\mathrm{o}$ and rearranging the previous equation:
\begin{equation}\label{eq:lead_lag}
	v_\mathrm{v} = \frac{1+R_\mathrm{v}C_\mathrm{v}s}{1+R_\mathrm{v}J_\mathrm{v}s}v_\mathrm{o}=G_\mathrm{vll}(s)v_\mathrm{o}.
\end{equation}  
The obtained expression represents the most general form of the VLL, which establishes a relationship between the virtual back EMF voltage and the measured output voltage. Based on this form, three different options for the VLL implementation can be derived: (i) lead-lag filter, (ii) PI controller, and (iii)~low-pass filter, which are described below.

\subsection{Lead-Lag Filter}
By considering the following parameterization of the obtained equation
\begin{equation}\label{eq:leadlag_vll}
	v_\mathrm{v} = \frac{1+R_\mathrm{v}C_\mathrm{v}s}{1+R_\mathrm{v}J_\mathrm{v}s}v_\mathrm{o}\implies v_\mathrm{v}=\frac{1+\frac{s}{\omega_\mathrm{lead}}}{1+\frac{s}{\omega_\mathrm{lag}}}v_\mathrm{o},
\end{equation} 
with
\begin{equation*}
     \omega_\mathrm{lead}=\frac{1}{R_\mathrm{v}C_\mathrm{v}}, \quad \omega_\mathrm{lag}=\frac{1}{R_\mathrm{v}J_\mathrm{v}},
\end{equation*}
it can be seen that the transfer function obtained is a lead-lag filter parameterized by frequencies $\omega_\mathrm{lead}$ and $\omega_\mathrm{lag}$. It is reasonable to assume that $J_\mathrm{v}>C_\mathrm{v}$ for any realistic machine and network parameters and hence $\omega_\mathrm{lead}>\omega_\mathrm{lag}$, implying a lag controller behavior. Lag controller is generally used to provide damping to the system. Nevertheless, the implementation is not restricted to this choice of parameters. 

\subsection{PI Controller}
\begin{figure}[!t]
	\centering
	\includegraphics[scale=1]{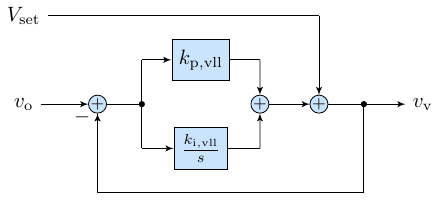}
	\caption{Voltage-locked loop control diagram.}
	\label{fig:vll_diagram}
\end{figure}

A voltage-locked loop can also be designed by considering the equivalence between a lead-lag filter and a PI controller. Namely, by selecting the PI controller gains as:
\begin{subequations}\label{eq:vll_gains}
    \begin{align}
	k_\mathrm{p,vll} &= (\frac{\omega_\mathrm{lead}}{\omega_\mathrm{lag}}-1)^{-1}=\frac{C_\mathrm{v}}{J_\mathrm{v}-C_\mathrm{v}},\\ k_\mathrm{i,vll}&= k_\mathrm{p,vll}\omega_\mathrm{lead}= \frac{1/R_\mathrm{v}}{J_\mathrm{v}-C_\mathrm{v}},
\end{align}
\end{subequations}

and implementing the PI controller with a negative feedback loop as indicated in Fig.~\ref{fig:vll_diagram}, an equivalent transfer function between the virtual back EMF and the output voltage as in \eqref{eq:lead_lag} is established. The voltage setpoint $V_\mathrm{set}$ corresponds to the desired output DC voltage value and is used here as a feedforward term.

Similarly to the PLL used in AC grid-connected converters, the obtained VLL unit aligns the internal (virtual EMF) voltage $v_\mathrm{v}$ with the output DC voltage $v_\mathrm{o}$. The speed (i.e., dynamics) at which the internal voltage tracks the output voltage is defined by the VLL gains \eqref{eq:vll_gains}.

\subsection{Low-pass Filter}
Simplified variants of \eqref{eq:lead_lag} can be derived by neglecting certain parts of the virtual DC machine. For instance, by neglecting the virtual capacitor, i.e., by setting $C_\mathrm{v}=0$, the VLL becomes
\begin{equation}\label{eq:lowpass_vll}
	v_\mathrm{v} = \frac{1}{1+R_\mathrm{v}J_\mathrm{v}s}v_\mathrm{o}\; \Rightarrow \; v_\mathrm{v}=\frac{1}{1+\frac{s}{\omega_\mathrm{lp}}}v_\mathrm{o}, \;\omega_\mathrm{lp}=\frac{1}{R_\mathrm{v}J_\mathrm{v}},
\end{equation}
where $\omega_\mathrm{lp}$ is the cut-off frequency of the obtained low-pass filter. This variant can be implemented using the VLL structure in Fig.~\ref{fig:vll_diagram} by setting 
\begin{equation}\label{eq:low-pass-gain}
	k_\mathrm{p,vll}=0,\quad k_\mathrm{i,vll}=1/R_\mathrm{v}J_\mathrm{v}.
\end{equation}

\section{Control Design} \label{sec:GFVCC_overview}
\begin{figure*}[!h]
	\centering
	\includegraphics[scale=0.77]{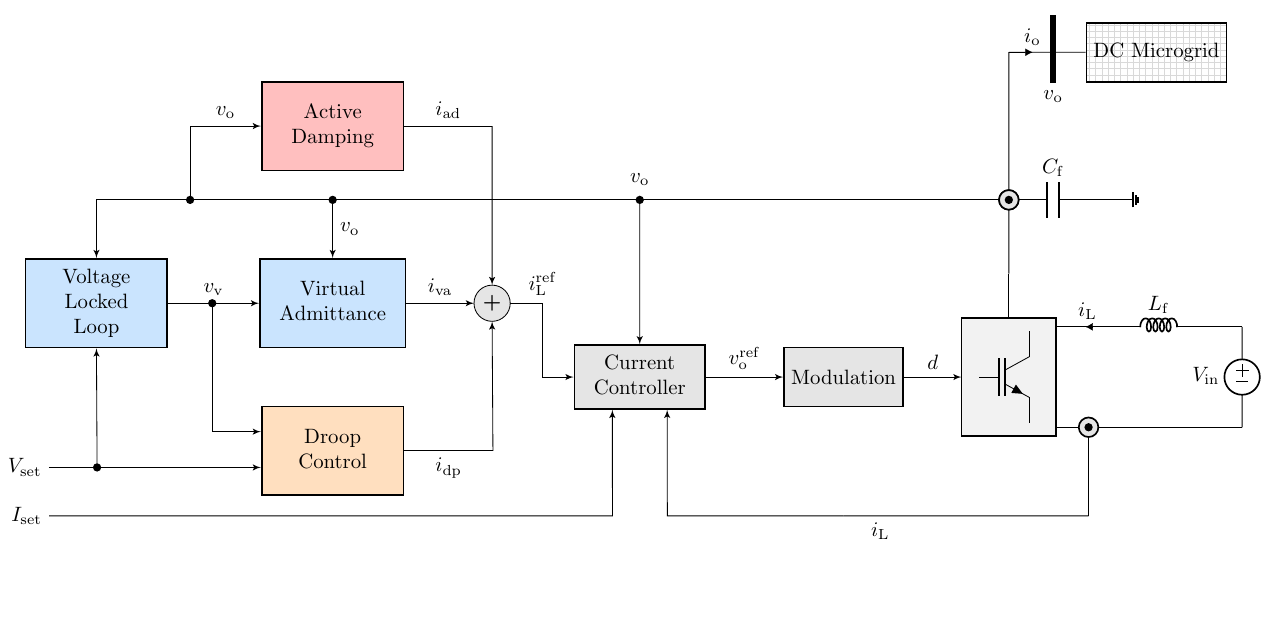}
    \vspace{-1cm}
	\caption{General configuration of the considered system.}
	\label{fig:sys_diag}
\end{figure*}
\subsection{System Configuration}
\begin{figure}[!b]
	\centering
	\includegraphics[scale=1]{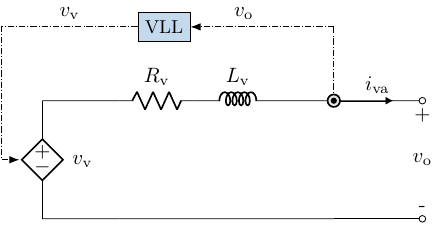}
	\caption{Circuit representation of the virtual admittance.}
	\label{fig:virt_admittance_circ}
\end{figure}
An overview of the considered system is depicted in Fig.~\ref{fig:sys_diag}, where an exemplary boost converter with input inductor $L_\mathrm{f}$ and output capacitor $C_\mathrm{f}$ is connected to a DC microgrid via a terminal DC bus. The converter is controlled using the VLL-IDC scheme, which is presented using control blocks in the figure. The control scheme comprises the standard current controller and a modulator (depicted in gray), as well as additional control features specific to VLL-IDC (depicted with colored blocks) proposed in this paper.

As already mentioned, the core principle behind VLL-IDC lies in emulating the combined effect of a virtual DC machine and a parallel-connected current source. The behavior of the virtual DC machine is emulated using a voltage-locked loop and a virtual admittance (both depicted in blue), while the virtual current source (depicted in orange) enables droop control and fast setpoint tracking. The droop control ensures power balancing in the steady-state, whereas the virtual DC machine is active only during transients, thereby improving the transient response. These features are derived and explained in the following subsections.

\subsection{Virtual Admittance}
Contrary to DC machine emulation concepts previously developed in the literature, the reformulations of the DC machine equations considered in this paper have led to the design of the VLL, rather than current or voltage references, to be applied at the converter terminals. To define the converter response to the output voltage deviations, we restructure the circuit in Fig.~\ref{fig:virt_DCm} by removing the DC machine and inserting the VLL instead. The current references are redefined and redirected towards the output voltage. Furthermore, the virtual capacitance is also removed in order to avoid derivative control action applied to the output voltage $v_\mathrm{o}$.
The obtained circuit is presented in Fig.~\ref{fig:virt_admittance_circ}, where the virtual DC machine is replaced by a controlled voltage source governed by the VLL. 

The virtual admittance of the armature circuit is obtained as the inverse of the virtual armature impedance, as:
\begin{equation}
	Y_\mathrm{v}=\frac{1}{R_\mathrm{v}+sL_\mathrm{v}}.
\end{equation}
The virtual armature current reference $i_\mathrm{va}$ can now be defined based on the presented circuit:
\begin{equation}\label{eq:i_va}
	i_\mathrm{va} = Y_\mathrm{v} (v_\mathrm{v}-v_\mathrm{o})=\frac{1}{R_\mathrm{v}+sL_\mathrm{v}}(v_\mathrm{v}-v_\mathrm{o}).
\end{equation}
The virtual admittance is ideally implemented with a full dynamic model, as described above, since its static counterpart reduces to merely emulating a virtual resistance $R_\mathrm{v}$.

Considering that the virtual EMF $v_\mathrm{v}$ is related to the output voltage $v_\mathrm{o}$, the virtual armature current can be directly related to the output voltage. Assuming that the VLL is implemented as a lead-lag controller \eqref{eq:lead_lag}, the expression can be derived as:
\begin{equation*}
	i_\mathrm{va} = \frac{1}{R_\mathrm{v}+sL_\mathrm{v}}(v_\mathrm{v}-v_\mathrm{o}) = \frac{1}{R_\mathrm{v}+sL_\mathrm{v}}(\frac{1+R_\mathrm{v}C_\mathrm{v}s}{1+R_\mathrm{v}J_\mathrm{v}s}v_\mathrm{o}-v_\mathrm{o}),
\end{equation*}
which, after rearranging, leads to
\begin{equation}\label{eq:i_va_direct}
	i_\mathrm{va} = \frac{(C_\mathrm{v}-J_\mathrm{v})s}{(1+L_\mathrm{v}/R_\mathrm{v}s)(1+R_\mathrm{v}J_\mathrm{v}s)}v_\mathrm{o}.
\end{equation}
Interestingly, the transfer function has one zero and two poles: one associated with the virtual admittance and one with the virtual inertia.
By choosing $J_\mathrm{v}=C_\mathrm{v}$, the response of the virtual admittance vanishes. This behavior arises since voltages at both ends of the virtual admittance would be perfectly synchronized as the virtual EMF and the output voltage would change with the same dynamics due to $J_\mathrm{v}=C_\mathrm{v}$.

On the other hand, assuming that the VLL is implemented as a low-pass filter, i.e., by neglecting the virtual capacitance, the expression becomes:
 \begin{equation}\label{eq:va_current_lowpass_vll}
 	i_\mathrm{va} = \frac{-J_\mathrm{v}s}{(1+L_\mathrm{v}/R_\mathrm{v}s)(1+R_\mathrm{v}J_\mathrm{v}s)}v_\mathrm{o}.
 \end{equation}
We can observe that the form of the controller does not change when the virtual capacitance is neglected, only the value of the term in the numerator. In summary, three different ways of implementing the virtual admittance are viable \eqref{eq:i_va}, \eqref{eq:i_va_direct}, and \eqref{eq:va_current_lowpass_vll}.
 
\subsection{Droop Control}
The virtual admittance response presented in the previous section provides only a transient response, since in steady-state $v_\mathrm{v}\mapsto v_\mathrm{o}$ in \eqref{eq:lead_lag} and therefore, $i_\mathrm{va}=0$ according to \eqref{eq:i_va}. In order to achieve power balancing in a steady-state, an additional current reference is computed, which can be seen as connecting a current source in parallel with the virtual admittance. The additional current reference imposes a droop control law in an effort to limit the output DC voltage deviations from the desired setpoint $V_\mathrm{set}$, as 
\begin{equation}\label{eq:i_droop}
	i_\mathrm{dp} = K_\mathrm{d} (V_\mathrm{set}-v_\mathrm{v}).
\end{equation}

Additional insights and further implementation possibilities can be uncovered by analyzing the small-signal behavior of \eqref{eq:i_droop}. Removing the constants related to a steady-state and considering \eqref{eq:lead_lag}, one obtains
\begin{equation}
	i_\mathrm{dp} = -K_\mathrm{d} v_\mathrm{v} = -K_\mathrm{d}\frac{1+R_\mathrm{v}C_\mathrm{v}s}{1+R_\mathrm{v}J_\mathrm{v}s}v_\mathrm{o},
\end{equation}
which calculates the droop response based on the measured output voltage.
We observe that by disregarding the virtual capacitance, i.e., by setting $C_\mathrm{v}=0$, and thus using the low-pass filter VLL implementation, the resulting droop control simplifies to 
\begin{equation}
	i_\mathrm{dp} = \frac{1}{1+R_\mathrm{v}J_\mathrm{v}s}K_\mathrm{d}(V_\mathrm{set}-v_\mathrm{o}).
\end{equation}
Therefore, the droop control current reference can be computed in different ways depending on the type of VLL used.

\subsection{Active Damping}
A virtual termination resistor can (optionally) also be included in the scheme for improved stability characteristics via an additional current reference:
\begin{equation}\label{eq:i_ad}
	{i}_{\mathrm{ad}} = -\frac{1}{R_\mathrm{ad}} {G}_\mathrm{bp}(s){v}_\mathrm{o}= -G_\mathrm{ad}(s){v}_\mathrm{o},
\end{equation}
where band-pass filtered output DC voltage measurement is applied on a virtual damping resistor $R_\mathrm{ad}$ to compute the reference. The band-pass filter, denoted by ${G}_\mathrm{bp}(s)$, ensures that only desired frequency components are used for damping, improving passivity in the mid- and high-frequency ranges. 

\subsection{Complete Current Reference}
The current references generated by the virtual admittance, droop control, and active damping are superimposed to create a joint current reference to be sent to the current controller:
\begin{equation}\label{eq:joint}
	i_\mathrm{ref} = i_\mathrm{va} + i_\mathrm{dp} + i_\mathrm{ad} + I_\mathrm{set},
\end{equation}
where $I_\mathrm{set}$ is the desired current output in steady-state.
The current setpoint can also be computed based on the desired output power $P_\mathrm{set}$, as $I_\mathrm{set} = P_\mathrm{set}/V_\mathrm{set}$. In this way, the current setpoint is directly fed to the current controller without filtering, therefore ensuring fast setpoint changes.

\begin{figure}[!b]
	\centering
	\includegraphics[scale=0.925]{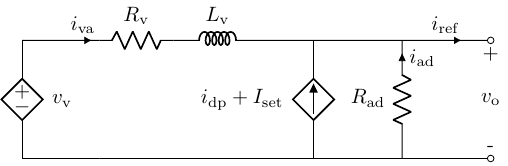}
	\caption{Circuit representation of the reference current generation.}
	\label{fig:complete_dcVLL-IDC_circuit}
\end{figure}

The generation of the current reference can be shown illustratively by the circuit in Fig.~\ref{fig:complete_dcVLL-IDC_circuit}, with the setpoint $I_\mathrm{set}$ assumed to be part of the virtual current source.

Further properties of the scheme can be uncovered by analyzing a few special cases. Considering that all current references are created only based on the output voltage measurement. A single-input-single-output transfer function between the output voltage and the joint current reference $i_\mathrm{ref}$ can be established. For simplicity, the active damping current reference is neglected as it is not a core part of the proposed scheme. The small-signal relationship is derived by considering the following steps:
\begin{align}
	i_\mathrm{ref} &= i_\mathrm{va} + i_\mathrm{dp}\\
	&= \frac{(C_\mathrm{v}-J_\mathrm{v})s}{(1+L_\mathrm{v}/R_\mathrm{v}s)(1+R_\mathrm{v}J_\mathrm{v}s)}v_\mathrm{o}-K_\mathrm{d}\frac{1+R_\mathrm{v}C_\mathrm{v}s}{1+R_\mathrm{v}J_\mathrm{v}s}v_\mathrm{o}.\nonumber
\end{align}
Virtual resistance can be selected as an inverse of the droop gain $K_\mathrm{d}=1/R_\mathrm{v}$ as a special case. Note that this choice might not be meaningful in all applications. Substituting this relationship into the previous equation, we obtain:
\begin{equation}
	i_\mathrm{ref} =-\frac{1}{R_\mathrm{v}}\left(1+\frac{(C_\mathrm{v}-J_\mathrm{v})L_\mathrm{v}s^2}{(1+L_\mathrm{v}/R_\mathrm{v}s)(1+R_\mathrm{v}J_\mathrm{v}s)}\right)v_\mathrm{o}.
\end{equation}
Two important observations can now be made: (i) the second term vanishes as $t\mapsto\infty$ by the final value theorem and thus only $R_\mathrm{v}$ impacts the steady-state,  (ii) there are only four parameters to be selected in this scheme $R_\mathrm{v}, J_\mathrm{v}, C_\mathrm{v}, L_\mathrm{v}$. 

The analysis above pertains to the lead-lag implementation of the VLL \eqref{eq:lead_lag}. By performing the same derivation steps, a similar relationship can be obtained for the case when the VLL is implemented as a low-pass filter \eqref{eq:lowpass_vll}, and is given by:
\begin{equation}
	i_\mathrm{ref} =-\frac{1}{R_\mathrm{v}}\left(1+\frac{L_\mathrm{v}/R_\mathrm{v}s}{(1+L_\mathrm{v}/R_\mathrm{v}s)(1+R_\mathrm{v}J_\mathrm{v}s)}\right)v_\mathrm{o}.
\end{equation}
Only three parameters need to be selected, $R_\mathrm{v}$ to define steady-state power sharing, and $L_\mathrm{v}, J_\mathrm{v}$ to define the dynamic response.

\section{Output Impedance Analysis and Parameter Tuning Guidelines}\label{sec:outputimpedance}
In this section, a small-signal model is developed to analyze the dynamic behavior of a bidirectional boost converter controlled by the proposed VLL-IDC method. Using this model, the closed-loop output impedance is derived. Based on the frequency response characteristics of the output impedance, the key regulation performance and stability properties of the proposed control scheme can be uncovered, and tuning guidelines can be established. 

\subsection{Small-Signal Model}
Starting from the small-signal equivalent circuit of the source converter in Fig.~\ref{fig:complete_dcVLL-IDC_circuit}, under the assumption the input voltage $V_\mathrm{in}$ remains constant, the inductor current $i_{
\mathrm{L}}$ and converter output voltage $ v_\mathrm{o}$ deviations can be expressed as linear functions of the duty-cycle $d$ and output current $i_\mathrm{o}$ perturbations, as follows:
\begin{align}
    i_{\mathrm{L}} &= G_\mathrm{di}(s) d
    +
    G_\mathrm{oi}(s)i_\mathrm{o}, \, \label{eq:ilf_general} \\
    v_\mathrm{o} &= G_\mathrm{dv}(s) d
    -
    Z_\mathrm{o}(s) i_\mathrm{o}, \,  \label{eq:vdc_general}
\end{align}
where $G_\mathrm{di}(s)$ and $G_\mathrm{oi}(s)$ represent the transfer functions from $d$ and $ i_\mathrm{o}$ to $i_{\mathrm{L}}$, respectively, while $G_\mathrm{dv}(s)$ denotes the transfer function from $d$ to $ v_\mathrm{o}$. In addition, $Z_\mathrm{o}(s)$ corresponds to the open-loop output impedance of the converter.

\begin{figure}[!t]
	\centering
    \includegraphics[width=\linewidth]{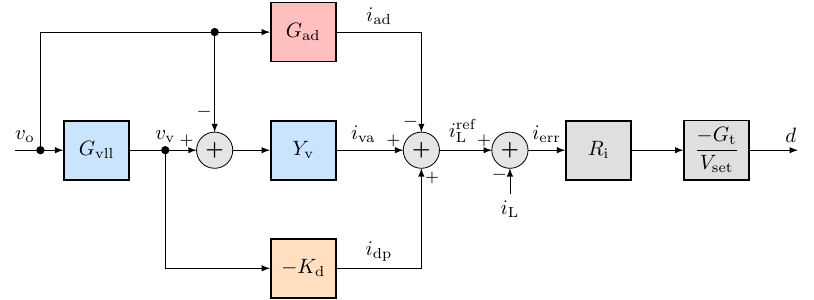}
	\caption{Linearized model of the VLL-IDC.}
	\label{fig:sys_lin}
\end{figure}

To derive the closed-loop output impedance, the converter’s control algorithm must be incorporated. For brevity, the explicit dependence on $s$ is omitted from the following derivations. 
Based on the small signal control block diagram from Figure~\ref{fig:sys_lin}, the control variable $d$ can be expressed in terms of the measurement signals $i_\mathrm{L}$ and $v_\mathrm{o}$, as:
\begin{equation}
    d = G_\mathrm{id}i_{\mathrm{L}} + G_\mathrm{vd}v_\mathrm{o} \, , \label{eq:duty_cycle}
\end{equation}
where $G_\mathrm{id} = G_\mathrm{t}R_\mathrm{i}/V_\mathrm{set}$ and
\begin{equation*}
    G_\mathrm{vd}=
    -\frac{G_\mathrm{t}R_\mathrm{i}}{V_\mathrm{set}}
    \left((G_\mathrm{vll}-1)Y_\mathrm{v}
    -G_\mathrm{vll}K_\mathrm{d}
    -G_\mathrm{ad}\right).
\end{equation*}
In these equations, $R_\mathrm{i}$ denotes the PI current controller, and $G_\mathrm{t}$ represents the delay introduced by PWM modulation, modeled as a second-order Padé approximation. Furthermore, $G_\mathrm{vll}$ is the VLL transfer function that maps $v_\mathrm{v}$ to $v_\mathrm{o}$ through the implemented VLL logic, while $Y_\mathrm{v}$ corresponds to the virtual admittance block used to generate the virtual admittance component of the current reference. Furthermore, $K_\mathrm{d}$ defines the voltage droop coefficient, which determines the droop component of the current reference, and $G_\mathrm{ad}$ denotes the active damping transfer function, as defined in equation~\eqref{eq:i_ad}.

With the derived duty-cycle function, we eliminate $i_{\mathrm{L}}$ by substituting~\eqref{eq:ilf_general} into~\eqref{eq:duty_cycle}:
\begin{equation}
    d = \frac{G_\mathrm{id} G_\mathrm{oi}}{1 - G_\mathrm{id} G_\mathrm{di}}  i_\mathrm{o} + \frac{G_\mathrm{vd}}{1 - G_\mathrm{id} G_\mathrm{di}} v_\mathrm{o} \, . \label{eq:duty_cycle_simplified}
\end{equation}
Finally, substituting~\eqref{eq:duty_cycle_simplified} into~\eqref{eq:vdc_general} and rearranging the terms leads to the expression for the closed-loop output impedance:
\begin{equation}
    Z_\mathrm{out} = -\frac{v_\mathrm{o}}{ i_\mathrm{o}} 
    = \frac{Z_\mathrm{o}(1 - G_\mathrm{id} G_\mathrm{di}) - G_\mathrm{dv}(G_\mathrm{id} G_\mathrm{oi})}
    {1 - G_\mathrm{id} G_\mathrm{di} - G_\mathrm{vd} G_\mathrm{dv}} \, . \label{eq:zout_generalized}
\end{equation}
%

\begin{table*}[b!]
    \centering
    \begin{equation}
        Z_\mathrm{out} = 
            \dfrac{s L_\text{f} + G_\text{t} R_\text{i}}
            {\Big( Y_{\mathrm{dc}} + \dfrac{D_\mathrm{set}^2}{s L_\text{f}} \Big)
            \big( s L_\text{f} + G_\text{t} R_\text{i} \big)
            + \dfrac{G_\text{t}R_\text{i}}{V_\mathrm{set}}
            \Big( I_\mathrm{set} - \dfrac{V_\mathrm{set} D_\mathrm{set}}{s L_\text{f}} \Big)
            \Big(
                D_\mathrm{set}
                + Y_\mathrm{v} (G_{\text{vll}} - 1) s L_\text{f}
                - G_\text{vll} K_\text{d} s L_\text{f}
                - {G_\text{ad} s L_\text{f}}
            \Big)}
    \label{eq:Zout}
    \tag{30}
    \end{equation}
\end{table*}

For the specific case of the bidirectional boost converter, the TFs in \eqref{eq:ilf_general} and \eqref{eq:vdc_general} can be written as:
\begin{subequations}
\begin{align}
    G_\mathrm{di}(s) &= -\frac{I_\mathrm{set}}{D_\mathrm{set}^2}\cdot\frac{1+\frac{Y_{\mathrm{dc}}V_\mathrm{set}}{I_\mathrm{set}}s}{1+\frac{L_\mathrm{f}Y_{\mathrm{dc}}}{D_\mathrm{set}^2}s} \, ,\\
    G_\mathrm{oi}(s) &= \phantom{-}\frac{1}{D_\mathrm{set}}\cdot\frac{1}{1+\frac{L_\mathrm{f}Y_{\mathrm{dc}}}{D_\mathrm{set}^2}s} \, ,\\
    G_\mathrm{dv}(s) &= -\frac{V_\mathrm{in}}{D_\mathrm{set}^2}\cdot\frac{1-\frac{L_\mathrm{f}I_\mathrm{set}}{V_\mathrm{in}D_\mathrm{set}}s}{1+\frac{L_\mathrm{f}Y_{\mathrm{dc}}}{D_\mathrm{set}^2}s} \, ,\\
    Z_\mathrm{o}(s) &= \phantom{-}\frac{1}{D_\mathrm{set}^2}\cdot\frac{L_\mathrm{f}s}{1+\frac{L_\mathrm{f}Y_{\mathrm{dc}}}{D_\mathrm{set}^2}s} \, .
\end{align}
\end{subequations}
Combined with the transfer functions previously derived from \eqref{eq:duty_cycle}, the output impedance is calculated using \eqref{eq:zout_generalized} and is given in \eqref{eq:Zout}, where $Y_{\mathrm{dc}}=1/(r_\mathrm{dc}+sC_\mathrm{dc})$ represents the admittance of the nonideal output capacitor.


\subsection{Control Tuning}
In the process of selecting control parameters, it is assumed that the electric circuit has already been properly designed and that the values of $L_\mathrm{f}$, $C_\mathrm{f}$, and the switching frequency $f_\mathrm{sw}$ are predetermined. In the following, we outline a systematic approach for the control parameter tuning of the VLL-IDC.

\textit{Current Controller:}
Given the system parameters, the PI current controller $R_\mathrm{i}(s)$ can be appropriately tuned. A standard internal model matching approach \cite{Harnefors1998} is employed, wherein the controller bandwidth is first selected, and the proportional and integral gains are subsequently derived based on the system parameters and the chosen bandwidth. Typically, the bandwidth of the current controller is set as $\omega_{\beta_\mathrm{i}} \cong 2\pi f_\mathrm{sw} / 10$.

\textit{Droop Gain:}
The key parameter in any droop control method is the droop gain. It is typically determined by the ratio of the maximum converter output current to the maximum permissible voltage deviation, expressed as $K_\mathrm{d} = I_\mathrm{max} / \Delta V_\mathrm{max}$. This selection ensures that the converter delivers the maximum allowable current when the voltage deviation reaches $\Delta V_\mathrm{max}$, as specified by relevant standards~\cite{currentos2025}.

\textit{Virtual Admittance:}
The virtual admittance comprises a resistance and an inductance. In droop control schemes, the virtual resistance is typically defined as the inverse of the droop gain, i.e., $R_\mathrm{v} = 1 / K_\mathrm{d}$. The virtual inductance, on the other hand, determines the dynamic response of the virtual admittance block. To achieve a faster response and improved stabilization, it is generally desirable to select a low value for the virtual inductance. However, the inductance value should not be excessively low, as this could compromise the ability of the current controller to track the reference signal generated by the virtual admittance block. Therefore, it is recommended to select the virtual inductance as $L_\mathrm{v} \cong 10 R_\mathrm{v} / \omega_{\beta_\mathrm{i}}$ to guarantee the desired time scale separation in the control loops.

\textit{Voltage-Locked Loop:} 
The tuning of the VLL can be approached in two ways. The first involves selecting the virtual moment of inertia $J_\mathrm{v}$ and virtual capacitance $C_\mathrm{v}$ based on physical intuition, typically within the ranges $J_\mathrm{v} \in [0.01,1]~\mathrm{kg{\cdot}m^2}$ and $C_\mathrm{v} \in [10, 50]C_\mathrm{dc}$. Note that $J_\mathrm{v}>C_\mathrm{v}$. The second approach relies on appropriately designing the lead-lag filter, which is equivalent to determining the PI controller parameters $k_\mathrm{p,vll}$ and $k_\mathrm{i,vll}$. In this case, a lower bandwidth yields slower response with lower $\mathrm{d}v_\mathrm{o}/\mathrm{d}t$ and more stable behavior, while a higher bandwidth enables a faster response at the expense of reduced robustness. These parameters strongly influence the shape of the output impedance and can therefore be utilized to achieve the desired impedance characteristics, as demonstrated in the following subsection.

\textit{Active Damping:}
The active damping current reference is tuned based on the \textit{passivity theorem}. As outlined in \cite{Lazarevic2025}, the goal is to maintain a strictly passive output impedance phase above a selected frequency threshold, typically a few hundred hertz. To achieve this, the band-pass filter ${G}_\mathrm{bp}(s)$ is designed to introduce a positive phase shift in the mid-to-high frequency range, while the virtual resistor $R_\mathrm{ad}$ is selected such that the phase requirement is maintained with a satisfactory magnitude response, i.e., avoiding magnitude peaks that could cause voltage overshoots, as discussed in \cite{Liu2020}.
\begin{figure}[t!]
	\centering
	\includegraphics[width=\linewidth]{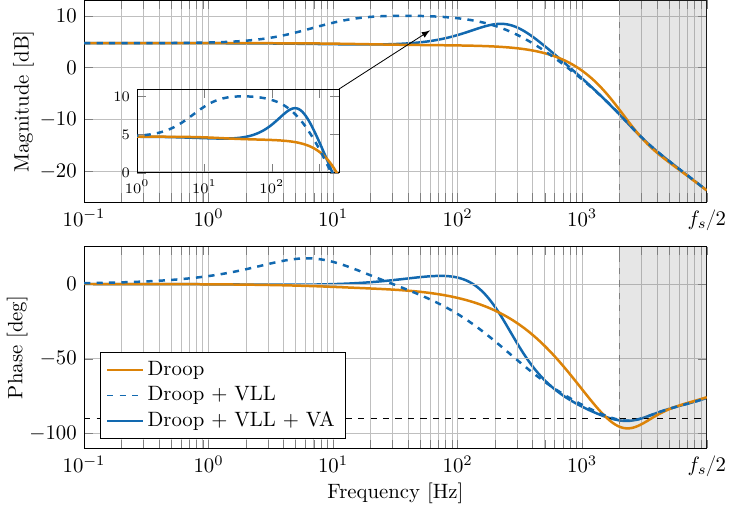}
	\caption{Bode diagram of the output impedance derived in \eqref{eq:Zout} for different cases: droop only (orange), droop with VLL (dashed blue), and droop with VLL and VA (solid blue). The shaded gray area indicates frequencies above the current control bandwidth.}
	\label{fig:output_impedance}
\end{figure}

\subsection{Output Impedance Analysis}
To further analyze the impact of different control parts of the scheme from Figs.~\ref{fig:sys_diag}~and~\ref{fig:sys_lin} on the resulting output impedance of the VLL-IDC control, let us observe the bode plots in Fig.~\ref{fig:output_impedance}. The transfer functions are parameterized according to Table~\ref{tab:VLL-IDC_exp_params}. The orange curve represents the basic droop control, which exhibits a non-passive high-frequency region that can compromise stability under certain operating conditions. To address this, the VLL can be incorporated in the droop path (dashed blue curve), which effectively eliminates the non-passive behavior. However, a relatively high impedance-magnitude overshoot occurs in the mid-frequency range, which can lead to large voltage excursions during transients \cite{Liu2020, DCGridForming2026}. Introducing the virtual admittance (VA) reduces the impedance magnitude in this region, as shown by the dashed blue curve, thereby improving dynamic performance. 

\begin{figure}[t!] 
	\centering
	\includegraphics[width=\linewidth]{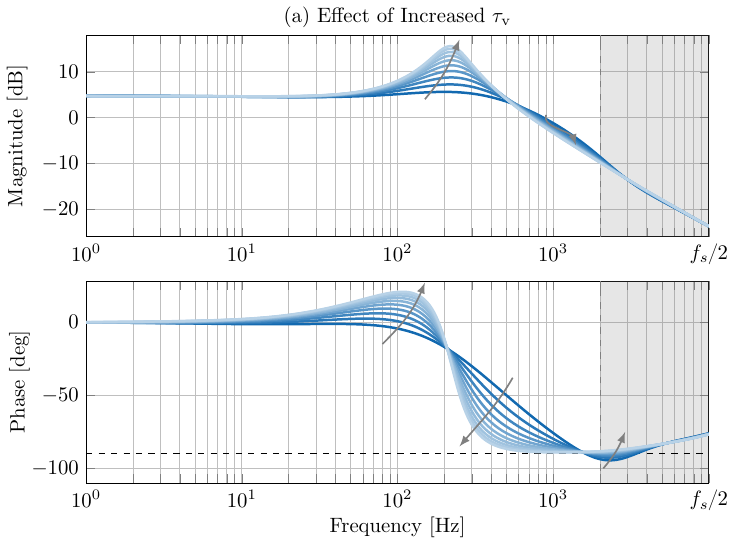}
	\includegraphics[width=\linewidth]{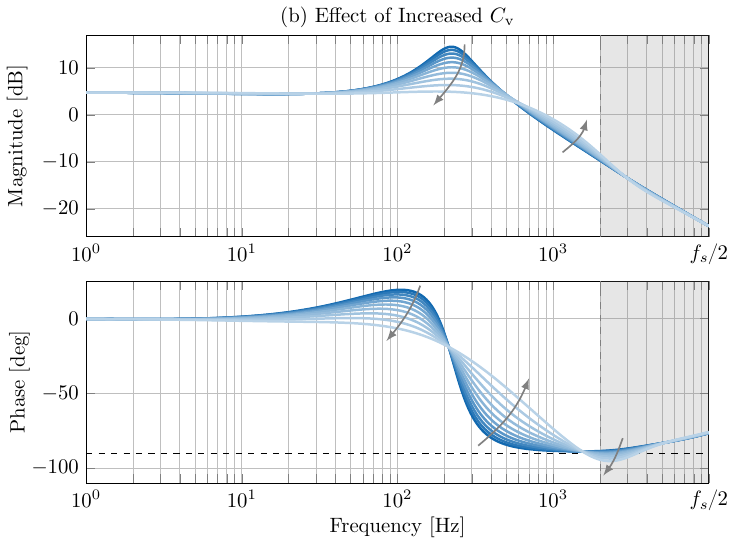}
	\caption{Bode diagram of the output impedance for increasing values of (a) $\tau_\mathrm{v}$ and (b) $C_\mathrm{v}$, where lighter shades correspond to larger parameter values.}
	\label{fig:Jv_Cv_sweep}
\end{figure}

In Fig.~\ref{fig:Jv_Cv_sweep}, the output impedance from \eqref{eq:Zout} is plotted for increasing values of the virtual inertia time constant $\tau_\mathrm{v}=R_\mathrm{v}J_\mathrm{v}$ and virtual capacitance $C_\mathrm{v}$, shown in subfigures (a) and (b), respectively. The virtual inertia $\tau_\mathrm{v}$ is analyzed instead of $J_\mathrm{v}$ due its more intuitive physical meaning in the control context. The value of $R_\mathrm{v}$ is kept constant during the parameter sweep of $\tau_\mathrm{v}$. The progression toward lighter shades indicates larger parameter values. As observed in Fig.~\ref{fig:Jv_Cv_sweep}(a), increasing $\tau_\mathrm{v}$ raises the output impedance magnitude across the mid-frequency range and increases the phase at both low and high frequencies. Conversely, Fig.~\ref{fig:Jv_Cv_sweep}(b) shows that increasing $C_\mathrm{v}$ reduces the magnitude overshoot in the mid-frequency region and increases the phase in the $10^2$--$10^3$~Hz range, improving damping and stability, respectively, but at the cost of reduced phase at higher frequencies and a potential loss of passivity.

\section{Experimental Results}\label{sec:exp}
\begin{figure}[!b]
	\centering
	\includegraphics[width=\linewidth]{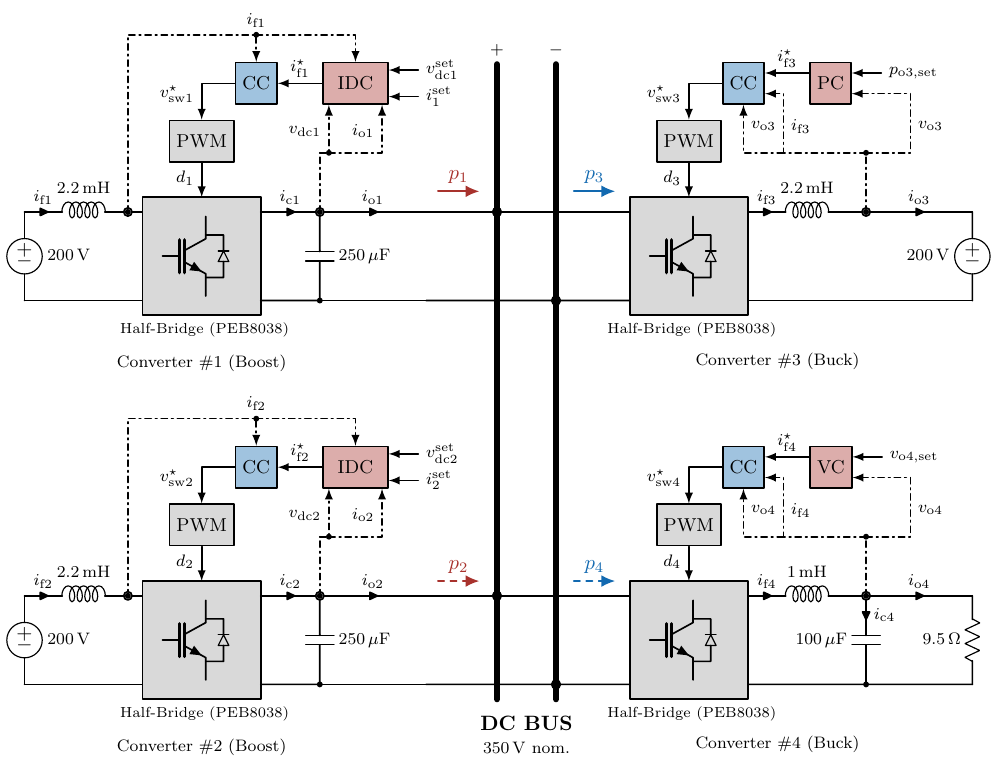}
	\caption{Four-converter experimental system where two boost converters on the left form the DC bus using the IDC (VLL-IDC) control; the power-controlled buck (top right) delivers into a 200V lab source, and the voltage-controlled buck (bottom right) supplies a resistive load.}
	\label{fig:VLL-IDC_exp_system}
\end{figure}

In order to demonstrate the working principles of VLL-IDC control and to compare it with the basic droop control, an experimental setup is constructed. Figure~\ref{fig:VLL-IDC_exp_system} shows a circuit diagram of the experimental system with the component values directly indicated in the figure. Filter parameters of the converters are appropriately selected to limit the peak current and to filter out the voltage ripple. Two boost converters fed from \SI{200}{\volt} laboratory sources stabilize the \SI{350}{\volt} bus, and each is controlled using the VLL-IDC block cascaded with the inner current controller and the modulator. The loads are the power-controlled buck delivering into the \SI{200}{\volt} battery and the voltage-controlled
buck regulating \SI{100}{\volt} into a \SI{9.5}{\ohm} resistor. The system is implemented on the Imperix hardware where each half-bridge is a \mbox{PEB\,8038} module, all controllers run on the B-Box\,4 controller at $f_\mathrm{s}=\SI{20}{\kilo\hertz}$, and the waveforms are Cockpit acquisitions at one sample per switching period over \SI{200}{\milli\second} windows. The employed lab supply is a triple output \mbox{PSB\,20920-40} EA (Tektronix) product. Table~\ref{tab:VLL-IDC_exp_params} collects the boost power-stage parameters and the VLL-IDC tuning used in the experiments. At the initial operating point each boost converter provides $1\,\si{\kilo\watt}$ to supply a total of $2\,\si{\kilo\watt}$ of buck-interfaced load. The DC bus voltage stabilizes at \SI{349.4}{\volt}.

\begin{table}[!ht]
	\centering
	\caption{Boost power-stage and VLL-IDC parameters of the experimental setup.}
	\label{tab:VLL-IDC_exp_params}
	\begin{tabular}{lcr}
		\toprule
		Quantity & Symbol & Value \\
		\midrule
		Input voltage                  & $V_\mathrm{in}$             & \SI{200}{\volt}         \\
		Bus voltage setpoint           & $V_\mathrm{set}$            & \SI{350}{\volt}         \\
		Filter inductance              & $L_\mathrm{f}$              & \SI{2.2}{\milli\henry}  \\
		DC-link capacitance            & $C_\mathrm{f}$              & \SI{250}{\micro\farad}  \\
		Switching frequency            & $f_\mathrm{s}$              & \SI{20}{\kilo\hertz}    \\
		Total loop delay               & $T_\mathrm{d}$              & \SI{62.5}{\micro\second}\\
		\midrule
		Droop gain                     & $K_\mathrm{d}$              & \SI{1}{\ampere\per\volt}\\
		Droop current setpoint         & $i^\mathrm{set}$            & \SI{5}{\ampere}         \\
		Virtual resistance             & $R_\mathrm{v}$              & \SI{1}{\ohm}            \\
		Virtual inductance             & $L_\mathrm{v}$              & \SI{1.26}{\milli\henry} \\
		Virtual capacitance            & $C_\mathrm{v}$              & \SI{15.9}{\milli\farad} \\
		Virtual inertia time constant  & $\tau_\mathrm{v}=R_\mathrm{v}J_\mathrm{v}$ & \SI{31.8}{\milli\second} \\
		\midrule
		VLL lag corner                 & $\omega_\mathrm{lag}$       & $2\pi\cdot\SI{5.0}{\hertz}$ \\
		VLL lead corner                & $\omega_\mathrm{lead}$      & $2\pi\cdot\SI{10.0}{\hertz}$ \\
		Virtual-admittance bandwidth   & $\omega_{\beta,Y}$          & $2\pi\cdot\SI{126}{\hertz}$ \\
		Current-loop bandwidth         & $\omega_{\beta,i}$          & $2\pi\cdot\SI{1400}{\hertz}$ \\
		\bottomrule
	\end{tabular}
\end{table}

\subsection{Load Increase}
\begin{figure}[!t]
	\centering
	\includegraphics[width=\linewidth]{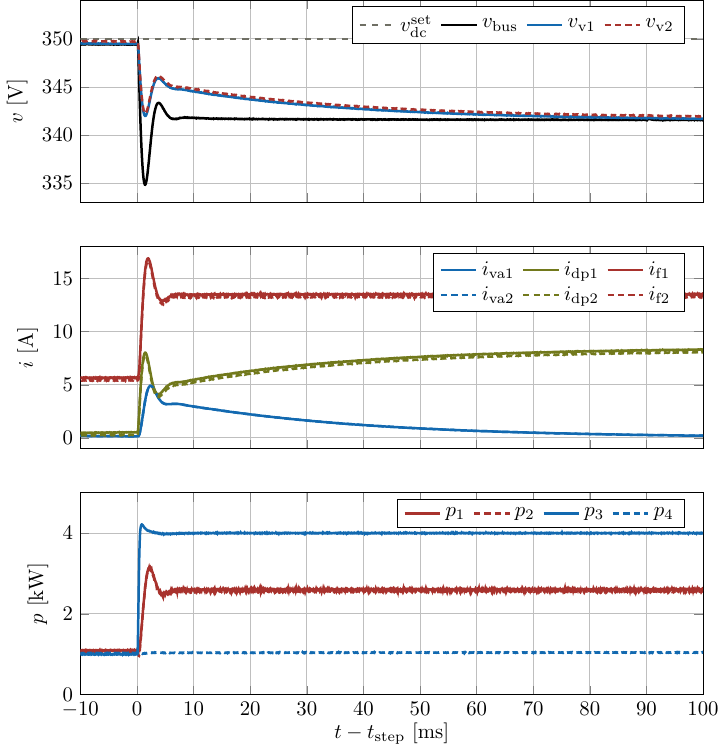}
	\caption{Recorded VLL-IDC response of the two boosts to the power step
		to $\SI{4}{\kilo\watt}$ of the power-controlled buck, showing, from top
		to bottom, the bus voltage and virtual back-EMFs, the internal currents
		(virtual admittance $i_\mathrm{va}$, droop $i_\mathrm{dp}$, and inductor
		$i_\mathrm{f}$), and the power outputs of all units in the system.}
	\label{fig:exp_VLL-IDC_pstep}
    \vspace{-0.35cm}
\end{figure}

Figure~\ref{fig:exp_VLL-IDC_pstep} shows the response of the two VLL-IDC-controlled boost converters to a load step, applied as a power increase of the power-controlled buck from \SI{1}{\kilo\watt} to \SI{4}{\kilo\watt}. Following the disturbance, the bus voltage first dips to \SI{334.8}{\volt} before it is recovered by the converters and settles at the droop level of \SI{341.5}{\volt} within \SI{6}{\milli\second}, exhibiting only a minor overshoot of \SI{1.8}{\volt} during the transient. The settled deviation of \SI{8.5}{\volt} is consistent with the shared droop current of \SI{8.5}{\ampere} per unit at $K_\mathrm{d}=\SI{1}{\ampere\per\volt}$, with the remainder of the \SI{13.5}{\ampere} unit current supplied by the constant \SI{5}{\ampere}  feed-forward.

The internal VLL-IDC variables reveal how this response is composed of two distinct timescales. As can be seen from the figure, the virtual back-EMFs $v_\mathrm{v1}$ and $v_\mathrm{v2}$ track the bus only through the VLL, and hence descend to \SI{341.8}{\volt} while the bus itself reaches a nadir nearly \SI{7}{\volt} lower. The resulting error is processed by the virtual admittances, whose currents spike to \SI{4.9}{\ampere} per unit and provide the fast response that arrests the dip within milliseconds. Correspondingly, the inductor currents peak near \SI{16.9}{\ampere} before easing to the \SI{13.5}{\ampere} unit value. Subsequently, a handover between the two functionalities takes place: as the VLL settles onto the new bus voltage, the virtual-admittance current $i_\mathrm{va}$ decays from \SI{3.2}{\ampere} to \SI{0}{\ampere} over roughly \SI{100}{\milli\second}, while the droop current $i_\mathrm{dp}$ rises from \SI{5.1}{\ampere} to \SI{8.5}{\ampere} in exact complement. Note that their sum, i.e., the current reference in \eqref{eq:joint}, together with the constant feed-forward remains fixed at \SI{13.5}{\ampere} throughout, so that the bus voltage is left undisturbed by the internal handover and the two units share within \SI{0.3}{\ampere} of each other at all times.

\subsection{Loss of a Grid-Forming Unit}
\begin{figure}[!t]
	\centering
	\includegraphics[width=\linewidth]{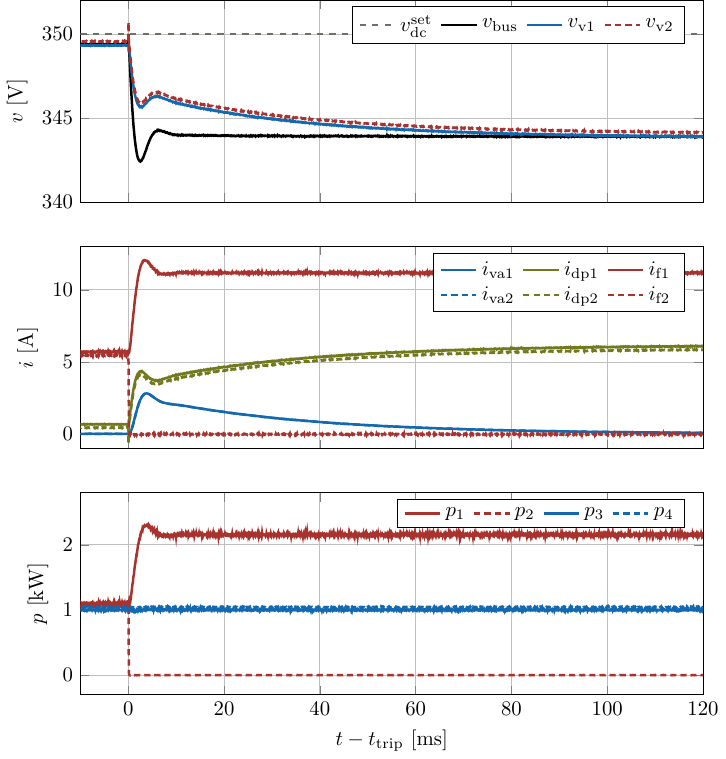}
	\caption{Recorded VLL-IDC response to the PWM-off disconnection of boost \#2 at the
		$(1+1)\,\si{\kilo\watt}$ dispatch, showing, from top to bottom, the bus voltage
		and virtual back-EMFs, the internal currents (virtual-admittance
		$i_\mathrm{va}$, droop $i_\mathrm{dp}$, and inductor $i_\mathrm{f}$), and the
		unit powers (the two boosts and the two bucks).}
	\label{fig:exp_VLL-IDC_trip}
    \vspace{-0.35cm}
\end{figure}

Figure~\ref{fig:exp_VLL-IDC_trip} illustrates a more severe event, namely the sudden
disconnection of boost \#2 through PWM blocking at the initial dispatch. Its inductor
current collapses to zero within a single control period, so that the surviving boost \#1
must instantly take over the entire generation duty. Despite this abrupt loss, the bus sags
by only \SI{7.0}{\volt} to \SI{342.4}{\volt} and returns to the new droop point of
\SI{343.9}{\volt} within \SI{3.5}{\milli\second}. As in the load step, the transient is
carried by the virtual admittance of the remaining unit, which injects a peak of
\SI{2.8}{\ampere} and then hands over to the droop as the VLL settles: the droop current
rises from \SI{0.7}{\ampere} to \SI{6.1}{\ampere} in complement with the decaying
virtual-admittance current, so that the surviving inductor current rises smoothly from
\SI{5.7}{\ampere} to \SI{11.2}{\ampere}. Note that the downstream load converters ride
through the event essentially unaffected.

Finally, it is worth emphasizing that, although the VLL operates on a filtered image of the DC bus voltage, the proposed scheme does not rely on an externally energized bus. It is therefore fully autonomous and capable of stand-alone operation. It is equally suited to black-starting a unenergized DC bus assuming a pre-charge circuit charges the bus to the minimum voltage in case the control scheme is implemented on a boost converter.

\subsection{Comparison with the Lead-Lag Droop}
\begin{figure}[!b]
    \vspace{-0.35cm}
	\centering
	\includegraphics[width=\linewidth]{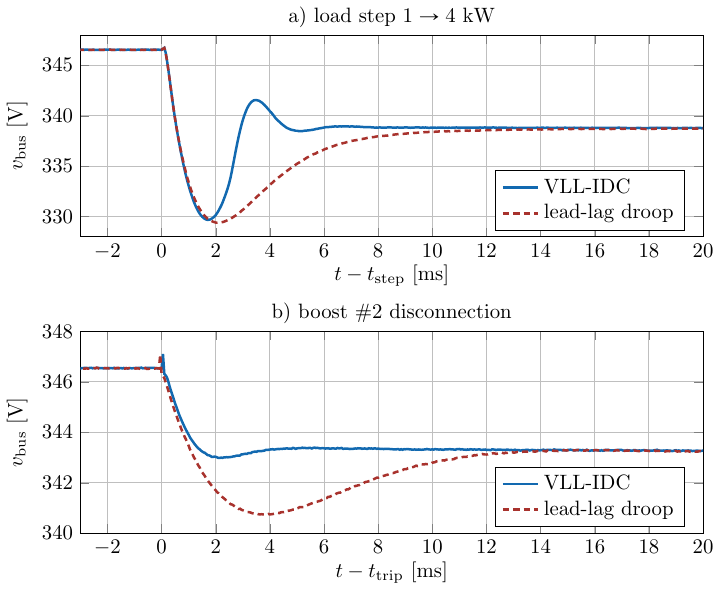}
	\caption{Recorded bus voltage under VLL-IDC and under the lead-lag droop for (a) the
		power step to $\SI{4}{\kilo\watt}$ of the power-controlled buck and (b) the
		disconnection of boost \#2, both recorded under the same operating conditions.}
	\label{fig:exp_VLL-IDC_vs_ll}
\end{figure}
Finally, Fig.~\ref{fig:exp_VLL-IDC_vs_ll} places the VLL-IDC directly against the lead-lag
droop, recommended by the emerging standards~\cite{currentos2025}, on the same two disturbances. To ensure a fair comparison,
both controllers are recorded at the same dispatch, with the same droop gain, and with
$I_\mathrm{set}=0$ in both cases. The lead-lag filter of the basic droop controller has a pole at \SI{20}{\hertz} and a zero at \SI{50}{\hertz}. As expected, the two schemes settle at identical
operating points, namely \SI{336.6}{\volt} after the load step and \SI{338.9}{\volt} after
the trip, since the steady state is dictated by the droop characteristic. The
distinction is therefore confined entirely to the transient, which is precisely the behavior
the VLL-IDC construction is designed to shape.

On the load step, the virtual admittance injects current early enough to arrest the
dip sooner, resulting in a smaller and earlier nadir of \SI{329.6}{\volt} against
\SI{326.9}{\volt} for the lead-lag droop. The difference in recovery is even more
pronounced: since the virtual admittance converts the voltage error into current at its full
\SI{126}{\hertz} bandwidth, the VLL-IDC bus returns to the droop level within
\SI{6}{\milli\second}, whereas the lead-lag droop approaches it only over roughly
\SI{10}{\milli\second}. A similar picture emerges for the disconnection event, where the early
support reduces the sag itself, from \SI{9.9}{\volt} to \SI{7.0}{\volt}, and shortens the
recovery from about \SI{10}{\milli\second} to \SI{3.5}{\milli\second}.

\section{Conclusion} \label{sec:concl}
This paper introduced VLL-IDC, an improved droop control for DC microgrids that separates the converter response into a fast virtual DC machine and a slow droop response from the virtual current source, synchronized to the bus through a voltage-locked loop, the DC counterpart of the grid-connected PLL.
The scheme has several practical advantages: it needs only the DC bus voltage measurement, has a simple linear structure with few parameters, works with any inner current controller, and lets the transient and steady-state support be tuned independently. The output-impedance analysis confirmed that the virtual admittance and VLL lower the mid-frequency impedance and restore passivity. On a DC microgrid setup, VLL-IDC delivered a markedly smaller voltage dip and faster recovery than the standard lead-lag droop under load steps, maintained accurate power sharing, and rode through the loss of a grid-forming unit with a minor bus sag and negligible disturbance to the downstream loads.

\bibliographystyle{IEEEtran}
\bibliography{bibliography}

\end{document}